%% file: neurips_2026.tex
\documentclass{article}

\usepackage[nonatbib,main,final]{neurips_2026}

\usepackage[utf8]{inputenc} %
\usepackage[T1]{fontenc}    %
\usepackage{hyperref}       %
\usepackage{url}            %
\usepackage{booktabs}       %
\usepackage{amsfonts}       %
\usepackage{nicefrac}       %
\usepackage{microtype}      %
\usepackage{xcolor}         %
\usepackage{wrapfig}
\usepackage[pdftex]{graphicx}
\usepackage{amsmath}
\usepackage{xspace}
\usepackage{multirow}

\usepackage[backend=biber,style=numeric-comp,sortcites=true,sorting=none]{biblatex}
\hypersetup{
    colorlinks,
    linkcolor={blue!64!black},
    citecolor={blue!64!black},
    urlcolor={blue!64!black}
}

\newcommand{\fettuccine}{FETTUCCINE\xspace}

\title{\fettuccine: Fast and efficient brain-to-text decoding on mobile devices}

\author{%
  Jonathan D. McCart$^{1,2,*}$ \enspace
  Pranav I. Deevi$^{1,2,*}$ \enspace
  Mehdi Azabou$^1$ \enspace
  Nanda H. Krishna$^{3,4}$ \\ \And
  Caleb A. McKinney$^{1}$ \enspace
  Nicholas S. Card$^{5}$ \enspace
  Sergey D. Stavisky$^{5}$ \enspace
  Chethan Pandarinath$^{1,2}$ \\[2pt]
  $^{1}$Georgia Tech \enspace
  $^{2}$Emory University \enspace
  $^{3}$Mila \enspace
  $^{4}$Universit\'e de Montr\'eal \enspace
  $^{5}$UC Davis \\[2pt]
  \href{https://fettuccine-brain2text.github.io}{\texttt{fettuccine-brain2text.github.io}}%
}

\begin{document}

\maketitle
{\renewcommand{\thefootnote}{\ensuremath{*}}%
 \footnotetext{Equal contribution. Correspondence to: \href{mailto:jmccart6@gatech.edu}{\texttt{jmccart6@gatech.edu}}, \href{mailto:pdeevi3@gatech.edu}{\texttt{pdeevi3@gatech.edu}}.}}

\begin{abstract}
Recent advances in speech neuroprostheses have enabled new communication avenues for people who have lost the ability to speak due to neurological illness or injury. These systems can decode neural activity during attempted speech into text by relying on language priors to achieve high decoding accuracy. This performance currently comes at the cost of using a compute-intensive language modeling stack, presenting a major bottleneck for real-world use. Further, current language models incur highly variable and disruptive inference latencies, limiting communication throughput and preventing users from reliably participating in natural conversation. We introduce \fettuccine, an end-to-end brain-to-text framework that leverages the speed and portability of automatic speech recognition (ASR) models to address the aforementioned practical challenges, while achieving usable ($<$10\%) word error rates (WER), efficient adaptation to neural non-stationarities, and reliable performance on mobile devices. When evaluated using publicly available data from an intracortical speech neuroprosthesis user, \fettuccine outperforms previous end-to-end methods, achieving as low as 4.39\% WER. Most notably, our models successfully run on a commercially-available mobile device with throughput reaching up to 370x real-time, providing the first demonstration of brain-to-text decoding on a mobile device. We also show that our models can be successfully finetuned to extend usable performance to future days. Beyond performance, our approach enables localization of the neural features most relevant for decoding via gradient-based saliency maps, which we show align with well-established physiological priors across different ASR model families. Taken together, our results show that \fettuccine overcomes core infrastructural and computational barriers, yielding a new class of portable brain-to-text communication.
\end{abstract}

\section{Introduction}

Speech neuroprostheses have demonstrated remarkable potential to restore communication in people with neurological illness or injury, capable of assisting users with amyotrophic lateral sclerosis (ALS) \cite{willett2023highperformance, Card2024SpeechNeuroprosthesis, jude2026decoding} and brainstem stroke \cite{Metzger2023, NTDeevi-pontine, Littlejohn2025StreamingBrainToVoice}, in some cases achieving accuracy above 99\% with vocabularies of over 125,000 words. Such systems already raise quality of life by helping people express themselves, reconnect with loved ones, and pursue active careers \cite{Card2025PersonalUse}. However, current systems have infrastructure requirements that are only tractable in research settings, prohibiting broad deployment. In particular, most current systems \cite{willett2023highperformance, Card2024SpeechNeuroprosthesis} use cascaded architectures: a neural decoder to predict phonemes, which are then fed into a separate stack of pretrained language models ($n$-gram, large language models (LLMs), etc.). These language models require dedicated local machines with $>$300 GB RAM together with a full consumer GPU, or persistent access to cloud servers. Such resource-intensive solutions limit a user's ability to use the system outside of research settings, or add network connectivity requirements and privacy risks when data leave the user's device for processing. 

A second challenge of current language modeling approaches is that they produce predictions with unreliable latency. When low-confidence or noisy phoneme predictions go through the language model stack, a search over the top $k$ most likely word sequences (i.e., beam search) expands exponentially, resulting in long delays before producing the final sentence. This sensitivity introduces substantial runtime variation across sentences, ranging from 52 ms to $>$10 seconds in standard approaches (see Figure \ref{fig:latency-raincloud}). Such variable delays can be frustrating or even dangerous in situations requiring fast and consistent communication (e.g., asking for help or physical assistance). More reliable latencies would allow users to better participate in high-bandwidth, naturally-paced conversations.

To help speech neuroprostheses translate into real-world applications, we sought to improve portability, reliability, and latency by lowering compute requirements and eliminating the cascaded language modeling stack. Our goal was to open a new tier of accessibility to speech neuroprosthetics technology: \textit{mobile devices}. Mobile on-device decoding would alleviate all of the above issues -- ensuring reliable access and usage, and preserving user privacy -- while substantially increasing the number of reachable users. When evaluating models for mobile deployment, we define a model's memory requirements as the total memory required to run inference (data loading, model weights, etc.). To evaluate these requirements, we consider one of the most ubiquitous commercial mobile devices, the Apple iPhone \cite{apple_iphone15promax_techspecs}, which uses a Unified Memory Architecture (UMA) that shares pages across the CPU, GPU, and Neural Engine \cite{apple_wwdc2024_10159}. Development standards suggest that models should be no larger than $\sim$1 GB \cite{coremltools}, with total inference processing using memory no larger than the available unified memory (e.g., for the iPhone used here, after OS and essential functions, typically $\sim$3 GB). Therefore, as a guide, we consider models suitable for mobile deployment if their total memory requirements for inference, after compression, are $\leq$3 GB. Such models are under the total UMA limit and would allow the use of other functions while decoding (e.g., music, web browsing, etc.). 

\begin{wrapfigure}[21]{r}{0.5\linewidth}
    \centering
    \vspace{-1.25em}
    \includegraphics[width=\linewidth]{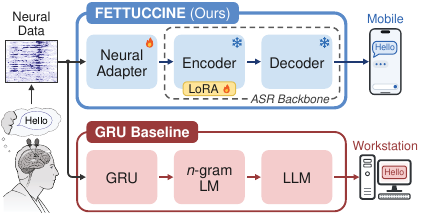}
      \caption{\textbf{Architecture and training strategy.} (Top) Our framework modifies open-source ASR models using a Neural Adapter network and LoRA updates targeted at the ASR encoder layers. Training occurs in two stages: a warm-up isolating the Neural Adapter, followed by joint training of the adapter and LoRA updates. The remaining ASR backbone remains frozen. (Bottom) A cascaded approach for comparison, featuring a non-end-to-end pipeline with a distinct GRU phoneme decoder.}
  \label{figure:architecture_and_training} 
\end{wrapfigure}

To achieve portable brain-to-text decoding, we leverage automatic speech recognition (ASR) models, which are often memory-efficient, streaming-ready, and commonly deployed on edge devices \cite{noroozi24_streamingconformer, 2026PushingLimitsOnDeviceStreaming}. Here we present \textbf{F}ast and \textbf{E}fficient \textbf{T}hought-to-\textbf{T}ext \textbf{U}nified \textbf{C}ross-modal \textbf{C}ommunication \textbf{I}nterfacing via \textbf{N}eural \textbf{E}mbeddings (\textbf{\fettuccine}), an ASR-enabled framework for portable brain-to-text communication. We show that a lean adaptation strategy (Figure \ref{figure:architecture_and_training}) generalizes across multiple ASR model families and enables state-of-the-art end-to-end decoding performance (see Table \ref{table:SOTA_E2E_WER}). Our pipeline also supports efficient finetuning for use on future days, and its differentiable structure enables novel methods to interpret neural data. Notably, \fettuccine can run on a mobile device (iPhone 15 Pro Max \cite{apple_iphone15promax_techspecs}) with decoding throughput up to 370x real-time. This achieves, to our knowledge, the first decoding of speech-related neural activity into text on an edge device -- an initial and important step towards untethering brain-to-text communication from research infrastructure.

\section{Related Work}
\paragraph{Cascaded approaches.} The standard brain-to-text pipeline \cite{willett2023highperformance, Card2024SpeechNeuroprosthesis, dietcorp, ryoo2026generalizable, tether_brain_whisperer, BIT} is a cascaded architecture: 1) neural data are decoded into phonemes using a neural network, 2) a Weighted Finite State Transducer (WFST)-based $n$-gram language model produces the top-$N$ candidate word sequences from the predicted phonemes, 3) a large language model rescores those candidates to refine the top predictions. The most memory-hungry step is phoneme-to-word modeling, where the $n$-gram language model (a fully-connected graph of transition probabilities) is loaded into RAM during inference. This graph is traversed via Viterbi Search to find the most likely word sequence from the predicted phoneme posteriors and previous words, requiring over 60 GB or 300 GB CPU RAM for the 3- or 5-gram language models, respectively. These memory requirements exceed typical consumer hardware and necessitate workstation or server hardware, which is even less practical for at-home or portable neuroprosthesis use. Additionally, the pipeline's latencies have an extremely long tail, with median per-sentence inference latency of 94 ms with worst case 15.9 s (Figure \ref{fig:latency-raincloud}).

\paragraph{End-to-end approaches.} Recent efforts \cite{feng24_interspeech, tether_brain_whisperer, BIT} demonstrated the feasibility of end-to-end decoding: translating neural activity directly into text in a single forward pass. By folding feature selection, word construction, and language modeling into one model, end-to-end approaches eliminate the language modeling stack that dominates the memory footprint of current systems. End-to-end methods have achieved competitive word error rates on the brain-to-text benchmarks \cite{Card2024SpeechNeuroprosthesis, b2txt2024lessons, brain_to_text_25}.

\paragraph{Concurrent work.} Two recent works, LightBeam \cite{LightBeam} and BrainWhisperer \cite{tether_brain_whisperer}, address related questions in brain-to-text decoding. We briefly contrast with these works below, and our performance is compared with their reported numbers in Table~\ref{table:SOTA_E2E_WER}. LightBeam \cite{LightBeam} is a cascaded approach that augments the standard language modeling stack by replacing the 5-gram WFST ($\sim$350~GB total memory) with a 4-gram language model based on KenLM \cite{KenLM} and an LLM with delayed fusion \cite{DelayedFusion}, achieving $\sim$15~GB total memory. While significant, this memory footprint is still prohibitive for mobile deployment. Additionally, LightBeam incurs a considerable $\sim$6x increase in latency over the GRU baseline. BrainWhisperer \cite{tether_brain_whisperer} demonstrates finetuning of OpenAI's Whisper \cite{Whisper} for brain-to-text. BrainWhisperer showed impressive word error rates of 8.7\% WER in end-to-end decoding, with relatively lean inference compute requirements (reported as $<$2 GB VRAM with 50~ms inference latency) \cite{tether_brain_whisperer}. They achieve this by adding adaptations such as day- and month-specific projections, Connectionist Temporal Classification \cite{CTCLoss} phoneme refinement, and phoneme-selective layers with local attention. In our work, we establish a framework that generalizes across different ASR families, providing evidence that ASR finetuning in general is a viable solution for portable brain-to-text. We accomplish this while also achieving lower word error rates (as low as 4.39\%, see Table \ref{table:SOTA_E2E_WER}) with a more straightforward modeling pipeline.

\section{Methods}

\paragraph{Datasets.} We use data from the brain-to-text 2025 challenge \cite{Card2024SpeechNeuroprosthesis}, which has microelectrode array recordings from the speech motor cortex of a clinical research participant, T15. The dataset includes the normalized threshold crossings and spiking-band power \cite{SamNasonSBP} from four 64-channel microelectrode arrays implanted in areas 4, v6v, d6v, and 55b \cite{Card2024SpeechNeuroprosthesis}. These data were recorded during a sentence copy task, where the participant is presented with a sentence that they then attempt to say. The cued text of each sentence serves as the labels for the dataset. The dataset contains 10,948 sentences from 45 research sessions spanning 20 months. Each session consists of several non-overlapping blocks of time in which sentences were cued to the participant. These blocks were randomly split into train, validation, and test sets. We separately use data from the brain-to-text 2024 challenge \cite{b2txt2024lessons}, a related dataset from a second participant (T12) performing the same sentence copy task. T12 has implants in the superior and inferior regions of area 44 and area 6v, and the dataset contains $\sim$10,000 sentences from 25 sessions spanning 4 months.

\paragraph{Architecture.} We build \fettuccine on current ASR architectures, which typically have 3 distinct components:  a convolution block, an encoder block, and a decoder block. The convolution block is meant to process audio features (typically mel-spectrogram) into a more abstract acoustic embedding space. In \fettuccine, we seek to exploit the potentially useful representations learned by the ASR models. We replace the initial convolution block with a Neural Adapter to directly map neural features into representations usable by the ASR model, which also allows us to control feature downsampling, enabling easy incorporation onto ASR architectures with different sampling rates.  Our Neural Adapter includes a temporal convolution variant of the U-Net \cite{UNet} and convolution sub-sampling \cite{rekesh23_fastconformer} (see Appendix \ref{appendix:front_end_deets} for details).  For the subsequent ASR model, we considered open-source models from the Whisper (Encoder-Decoder Transformer) \cite{Whisper} and Parakeet (Conformer with RNN-T) \cite{Parakeet, rekesh23_fastconformer, noroozi24_streamingconformer} families available from Hugging Face \cite{huggingface}. We use the tokenization schemes that are native to each ASR variant, where Whisper variants use the BPE tokenizer \cite{BPE_ref_old__sennrich2016neuralmachinetranslationrare, BPE_newer__radford2019language, Whisper} while the Parakeet variants use an adaptation of the SentencePiece tokenizer \cite{kudo2018sentencepiecesimplelanguageindependent}. To further increase the compatibility of the pretrained ASR model with neural activity, we optionally apply Low-Rank Adaptation (LoRA) updates \cite{lora} using parameter-efficient finetuning (PEFT) \cite{peft} on a subset of the backbone encoder layers. These design choices are justified in Appendix \ref{appendix:ablations}, where we show how each of these simple additions improves model performance.

\paragraph{Training strategy.} \fettuccine models are trained in two stages: Neural Adapter warmup, followed by the addition of LoRA. For each ASR backbone, we train \fettuccine models with the native loss used for that particular ASR backbone: either cross entropy (Whisper family), transducer loss \cite{Parakeet}, or the RNN-T loss \cite{RNN_T_Loss} (Parakeet EOU). We use a variety of augmentations (described below) throughout training to improve performance and robustness. All models are optimized with AdamW \cite{AdamW}, and hyperparameter settings are chosen by iterating towards optimal word error rate (WER) on the validation split. Our final settings for each \fettuccine model are given in Appendix \ref{appendix: HPs_fettuccine_whisper} and Appendix \ref{appendix: HPs_fettuccine_parakeet}. Details on training hardware are given in Appendix \ref{appendix:compute_resources}.

\paragraph{Neural augmentations.} Neural recordings exhibit many nonstationarities across sessions \cite{perge2013intraday, downey2018intracortical, sussillo2016robust, karpowicz2024falcon, karpowicz2025stabilizing}. To build invariance to such nuisance variability, we train models using augmentations from prior work \cite{Card2024SpeechNeuroprosthesis}, including per-channel augmentations (additive shifting, masking \cite{park2019specaugment}, and dropout), and global augmentations (temporal jitter, additive Gaussian white noise and smoothing across all neural features). Relevant hyperparameters are in Appendix \ref{appendix:augmentations}. We also designed new augmentations based on word-level variations in speech. Significant speech variation occurs naturally at the word level, as speakers often stretch, compress, emphasize, or soften individual words. Varying the amplitude and duration of the neural activity tied to each word might allow us to simulate such variation directly, thus increasing diversity and variability in the training data. However, introducing synthetic word-level augmentations requires putative word boundaries. We estimate these directly from the z-scored neural features by computing a neural envelope, which is a smoothed and filtered sum across both threshold crossings and spiking-band power. The neural envelope reveals periodic structure in neural activity, with the number of peaks often consistent with the number of words in a sentence (see Figure \ref{figure:interpretability}D for a representative example). Prior evidence shows that both threshold crossings and spiking-band power during speech correlate with speech envelope \cite{wairagkar2025instantaneous}. We therefore treat the neural envelope as a data-driven proxy for word timings. We computed estimated word-level sub-windows of the sentence using intervals whose lengths are determined by the duration of TTS-generated audio \cite{pyttsx3} for the words in the cued sentence. To find the timings within the sentence we used a dynamic programming approach that maximizes the aggregate neural activity within the intervals. Though inexact, we hypothesized that these boundaries could serve as reasonable pseudo-labels for three augmentations: \textit{word warping}, which time warps each word and its trailing silence independently; \textit{word gain}, which scales activity within each word region by an independent scaling factor, motivated by the loudness and vigor modulation observed in the ventral precentral gyrus~\cite{Srinivasan2026}; and \textit{word cropping}, which drops a random contiguous subset of words and the corresponding transcript tokens.

\section{Results}
\subsection{\fettuccine is the leanest, most performant end-to-end pipeline for brain-to-text \label{results:SOTA_E2E_WER}}
\begin{wrapfigure}[17]{r}{0.4\linewidth}
    \centering
    \vspace{-1.2em}
    \includegraphics[width=\linewidth]{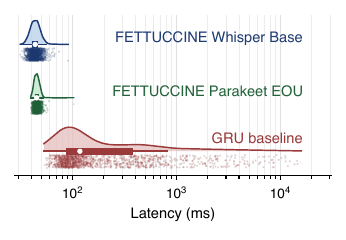}
    \vspace{-2em}
    \caption{{\bf Inference latency distributions.} We show the Whisper Base and Parakeet EOU variants of \fettuccine, alongside the GRU baseline \cite{Card2024SpeechNeuroprosthesis}. Each point represents a sentence from the brain-to-text 2025 validation set.}
    \label{fig:latency-raincloud}
\end{wrapfigure}
We evaluate \fettuccine on the brain-to-text 2025 test set ($n=1450$ sentences), which the competition splits into public and private subsets.  All experiments in this section reflect inference on server-grade NVIDIA hardware (A40). As shown in Table \ref{table:SOTA_E2E_WER}, all \fettuccine models achieve usable WER, with the best models outperforming prior end-to-end brain-to-text approaches. Against the GRU baseline and other cascaded approaches, \fettuccine models achieve competitive WER with a fraction of the memory and substantially more consistent latencies (see Figure \ref{fig:latency-raincloud}). Memory and latency numbers for the GRU baseline are representative of any cascaded approach that uses the same 5-gram model with LLM rescoring \cite{willett2023highperformance, Card2024SpeechNeuroprosthesis, dietcorp, ryoo2026generalizable, tether_brain_whisperer, BIT}. Table \ref{table:SOTA_E2E_WER} quantifies the throughput values in terms of inverse real-time factor (RTFx), which measures the speed of inference as a ratio of the duration of the sentence to the duration of model inference. We show RTFx for each \fettuccine variant, the GRU baseline and reported numbers for other relevant models.

Comparing performance and efficiency across ASR model families, we see that \fettuccine supports a range of distinct accuracy-latency profiles, allowing variant selection based on deployment priorities. Whisper variants, based on the encoder-decoder Transformer architecture \cite{Vaswani2017Transformers}, achieve the lowest WERs. Parakeet variants achieve comparable performance, but when accelerated by TensorRT \cite{tensorrt} are capable of substantially higher throughput on NVIDIA hardware (1379 RTFx). Applying standard inference optimizations to reduce latency (e.g., static KV caching, torch compile, etc.) allows Whisper variants to achieve at least 185 RTFx. End-to-end decoding combined with these optimizations produces \fettuccine models with consistently lower latency than the GRU baseline, whose WFST-based decoding has high latency variance (494.8 $\pm$ 2457.7 ms, with worst-case sentences exceeding 10 seconds). We also tested variants with beam search ($k=5$; see Appendix \ref{appendix:beam}) and found it consistently improved WER across all \fettuccine variants at approximately 2x latency cost, which may be particularly useful for applications that prioritize accuracy over throughput. These results show that the \fettuccine framework generalizes across ASR architectures trained with different loss functions, and that competitive end-to-end performance is achievable with a minimal adaptation strategy. They also show that \fettuccine can achieve remarkably fast and consistent latencies on NVIDIA hardware using publicly available acceleration strategies \cite{tensorrt}. 

To see whether \fettuccine generalizes to another participant, we include preliminary results for the brain-to-text 2024 dataset \cite{b2txt2024lessons}. We find that our best models continue to outperform BIT \cite{BIT}. Although our approach lags the cascaded approaches for this dataset, with further hyperparameter tuning we expect similar trends to those observed on the brain-to-text 2025 dataset.

\begin{table}[t]
  \caption{Accuracy and compute efficiency on brain-to-text 2024 (T12, $n = 1200$) and 2025 (T15, $n = 1450$). Our models were run on an NVIDIA A40 (batch size 1, greedy decoding); compute columns represent the worst case across participants (Table~\ref{tab:compute_T12_T15}). \textbf{Bold}: best; \underline{underline}: second best.}
  \label{table:SOTA_E2E_WER}
  \centering
  \small
  \setlength{\tabcolsep}{4pt}
  \newcommand{\ens}[1]{\textcolor{gray}{#1}}
  \resizebox{\textwidth}{!}{%
  \begin{tabular}{lcccccccc}
    \toprule
    Model & \multicolumn{2}{c}{T15 WER \%} & T12 WER \% & Params & Latency & RTFx & VRAM & RAM \\
    \cmidrule(lr){2-3}
          & pub $\downarrow$ & priv $\downarrow$ & test $\downarrow$ & B $\downarrow$ & ms $\downarrow$ & $\uparrow$ & MB $\downarrow$ & MB $\downarrow$ \\
    \midrule
    \multicolumn{9}{l}{\textbf{End-to-End \textit{(E2E)}}} \\
    \cmidrule(lr){1-9}
    \multicolumn{9}{l}{\textit{E2E (Ours) -- Greedy}} \\
    \quad Whisper Tiny       & 7.70          & 7.61          & 19.88              & \textbf{0.04}      & 37.9 $\pm$ 5.6           & 176           & 856           & \textbf{2006} \\
    \quad Whisper Base       & 6.61          & 6.72          & 15.56              & \underline{0.08}   & 50.2 $\pm$ 16.2          & 124           & \underline{700} & \underline{2040} \\
    \quad Whisper Small      & \textbf{4.39} & 5.05       & 15.98              & 0.25               & 98.2 $\pm$ 7.8           & 61            & 1594          & 2196 \\
    \quad Whisper Large v3 Turbo     & 4.64          & \underline{4.91} & 12.98      & 0.82               & 79.9 $\pm$ 24.7          & 77            & 2580          & 4914 \\
    \quad Parakeet TDT       & 6.39          & 6.34          & 11.04              & 0.63               & 48.9 $\pm$ 5.7           & 122           & 1920          & 2454 \\
    \quad Parakeet EOU       & 10.33         & 9.56          & 16.12              & 0.12               & 43.2 $\pm$ 6.7           & 139           & 866           & 2411 \\
    \quad Parakeet TDT-TRT   & 6.39          & 6.37          & 11.81              & 0.63               & \textbf{12.9} $\pm$ 1.9  & \textbf{577}  & 1610          & 4660 \\
    \quad Parakeet EOU-TRT   & 10.18         & 9.73          & 16.19              & 0.13               & \underline{19.6} $\pm$ 4.8 & \underline{447} & \textbf{644} & 2449 \\
    \cmidrule(lr){1-9}
    \multicolumn{9}{l}{\textit{E2E (Baselines)}} \\
    \quad BIT E2E (Aero1-Audio 1.5B) \cite{BIT}          & 11.06     & --       & 15.67     & $>$1.50   & $\sim$950 & --   & --           & --   \\
    \quad BrainWhisperer E2E (cross) \cite{tether_brain_whisperer}  & 8.70     & --       & --        & $>$0.24   & 50        & --   & $\leq$2000   & --   \\
    \midrule
    \multicolumn{9}{l}{\textbf{Cascaded approaches}} \\
    \cmidrule(lr){1-9}
    \quad GRU + 5-gram + OPT-6.7B \cite{Card2024SpeechNeuroprosthesis}                          & 6.67            & 7.00            & 9.76             & $>$6.7 & 494.8 $\pm$ 2457.7  & 162.76      & 18{,}100   & 320{,}000 \\
    \quad LightBeam (GRU) + 4-gram + Llama-3.2-1B  \cite{LightBeam}                                  & 5.77 & 6.47 & \underline{9.37} & $>$1.2 & --                  & $\sim$16.6  & 5800       & 11{,}600  \\
    \quad LightBeam (Transformer) + 4-gram + Llama-3.2-1B  \cite{LightBeam}                          & \underline{4.44} & \textbf{4.59} & \textbf{8.08} & $>$1.2 & --                  & $\sim$20    & --         & --        \\
    \bottomrule
  \end{tabular}}
  \par\vspace{2pt}
  \vspace{-1.25em}
\end{table}
\vspace{-0.2em}
\subsection{\fettuccine enables brain-to-text decoding on mobile devices \label{results:mobile_devices}}
\begin{wrapfigure}[15]{r}{0.4\linewidth}
    \vspace{-1.5em}
    \centering
    \includegraphics[width=0.745\linewidth]{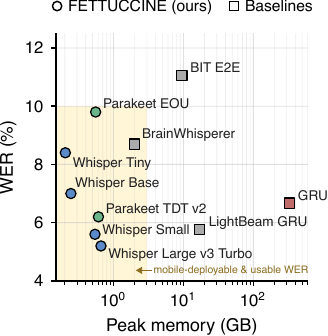}
    \vspace{-1mm}
    \caption{{\bf Brain-to-text 2025 val WER vs. Peak Memory.} Shaded: mobile-ready models (usable and portable).}
    \label{figure:wer-vs-memory}
\end{wrapfigure}
The core motivation of \fettuccine~-- enabling portable brain-to-text decoding -- requires that models fit on a mobile device and perform well under the associated resource constraints. We therefore evaluate models on both memory and WER. Figure \ref{figure:wer-vs-memory} demonstrates where models fall based on their performance and memory efficiency. The shaded region corresponds to models that would be usable ($<$10\% WER) and sufficiently portable ($<$3 GB peak memory). With the exception of concurrent work \cite{tether_brain_whisperer}, \fettuccine produces the only models that satisfy both criteria, with some variants requiring $<$1 GB of total memory during inference. The \fettuccine models in Figure \ref{figure:wer-vs-memory} reflect our on-device deployment sizes, while for baselines we consider either their publicly reported numbers, or their benchmarked performance and efficiency on supporting hardware. The GRU baseline was benchmarked on a server (NVIDIA A40 GPU, 512 GB RAM).

As empirical validation, we present the first demonstration of mobile brain-to-text decoding, using an iPhone 15 Pro Max \cite{apple_iphone15promax_techspecs}. As shown in Table \ref{table:mobile_devices}, all \fettuccine variants maintain usable word error rates and latencies while staying within mobile device constraints. To enable models to run on the iPhone, we convert all trained models from PyTorch into Core ML \cite{coremltools} and run inference in Swift. Word error rates closely match our results on NVIDIA hardware, while latencies exhibit minor differences, as expected when moving to a commercial mobile device. In the case of the large models (Whisper Large v3 Turbo and Parakeet TDT) we bisect each model's encoder into two chunks and run these sequentially, harming their latency but enabling mobile deployment (see Appendix \ref{appendix: mobile_decoding_details} for additional details). Optimizations for mobile deployment vary by ASR model family and model size, with further details explained in Appendix \ref{appendix: mobile_decoding_details}.

\begin{table}[h]
  \caption{Mobile device decoding and benchmarking. Shown are the sentence-wise mean and standard error for the brain-to-text 2025 validation set WER, latency, throughput (RTFx), and peak UMA (MB). * indicates that a subset of the model was bisected and run sequentially to fit on device.}
  \label{table:mobile_devices}
  \centering
    \begin{tabular}{lcccc}
    \toprule
    FETTUCCINE variant & WER (\%) ($\downarrow$) & Latency (ms) ($\downarrow$) & RTFx ($\uparrow$) & Peak UMA (MB) \\
    \midrule
    Whisper Tiny & 8.4 $\pm$ 0.4 & \textbf{48.4} $\pm$ 0.1 & \textbf{372.7} $\pm$ 2.7 & \textbf{209} \\
    Whisper Base & 7.0 $\pm$ 0.3 & 93.0 $\pm$ 0.2 & 194.3 $\pm$ 1.4 & \underline{251} \\
    Whisper Small & \underline{5.6} $\pm$ 0.3 & 328.4 $\pm$ 0.8 & 55.1 $\pm$ 0.4 & 555 \\
    Whisper Large v3 Turbo & \textbf{5.2} $\pm$ 0.3 & 2959.8 $\pm$ 4.6 & 6.2 $\pm$ 0.1 & 676* \\
    Parakeet EOU & 9.8 $\pm$ 0.4 & \underline{70.4} $\pm$ 0.5 & \underline{254.6} $\pm$ 0.7 & 567 \\
    Parakeet TDT v2 & 6.2 $\pm$ 0.3 & 619.7 $\pm$ 1.2 & 29.1 $\pm$ 0.2 & 622* \\
    \bottomrule
    \end{tabular}
\end{table}

\subsection{\fettuccine can be efficiently finetuned for new sessions and scales with pretraining} \label{scaling}

Given the challenges surrounding neural recording drift over time \cite{perge2013intraday, downey2018intracortical, sussillo2016robust, karpowicz2024falcon, karpowicz2025stabilizing}, a key practical question for brain-to-text models is whether they can generalize to new recording sessions. Thus, to study \fettuccine's data requirements and adaptability to future sessions, we consider a 39-session subset of the brain-to-text 2025 challenge dataset \cite{Card2024SpeechNeuroprosthesis} spanning $\sim$11.5 months. To ensure our evaluation setup mirrored practical deployment scenarios, we create a causal data split: our training set contains 34 sessions from August 2023 through April 2024, and our evaluation set contains 5 sessions from May through July 2024. We specifically evaluate the Whisper Base and Parakeet EOU variants of \fettuccine as they were among the most portable for each backbone architecture.

\begin{figure}[htbp]
    \centering
    \includegraphics[width=\linewidth]{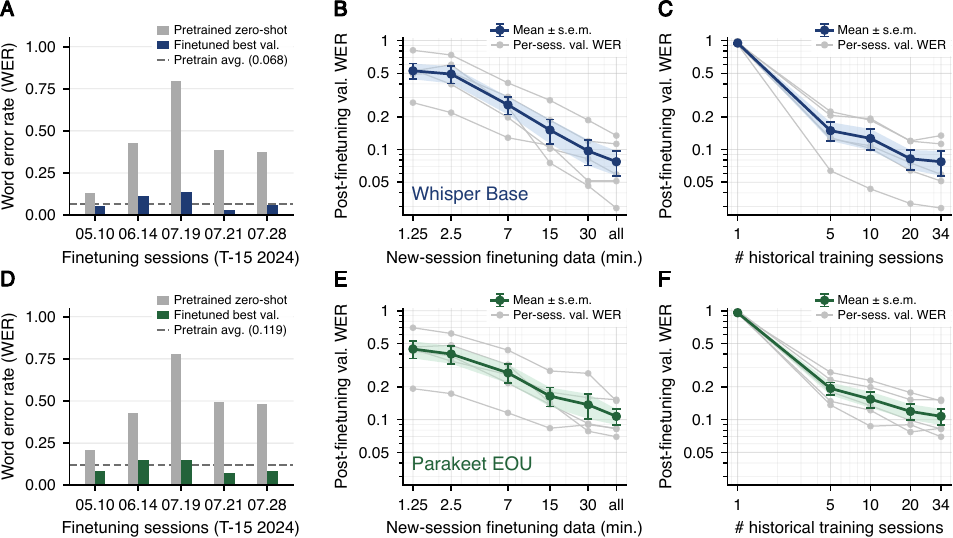}
    \caption{{\bf \fettuccine generalizes to held-out sessions and improves with more pretraining.} \textbf{(A)} \fettuccine with a Whisper Base backbone generalizes well to future sessions held-out from pretraining. \textbf{(B)} Pretrained models adapt well to held-out sessions in the low-data regime, needing just 15 minutes of finetuning data on average (and sometimes less) to achieve reasonable word error rates. `All' denotes all training data from the held-out session, which can range from 30 to 60 minutes. \textbf{(C)} Scaling the pretraining dataset size leads to improved finetuning performance on held-out future sessions. \textbf{(D--F)} Same as (A--C) but for \fettuccine with a Parakeet EOU backbone.}
    \label{figure:wer_ft_scaling}
\end{figure}

First, we evaluate \fettuccine's zero-shot and finetuning performance on the validation set from held-out sessions (Figure \ref{figure:wer_ft_scaling}A,D). As expected from prior work \cite{Metzger2023,karpowicz2024falcon}, \fettuccine does not generalize zero-shot to held-out sessions. The standard deployment practice for brain-to-text models is therefore to finetune them on data collected from the target session \cite{willett2023highperformance,Card2024SpeechNeuroprosthesis}. We find that finetuning on training data from the target session (even for just a few epochs; Table \ref{tab:finetuning-epochs}) rescues performance to usable WERs, indicating that our pretrained models are highly amenable to session-specific adaptation. Furthermore, we find that models finetuned for the immediate previous session substantially reduce the zero-shot WER on held-out sessions compared to the pretrained model (Table \ref{tab:prev-day-zs}). This is particularly useful to practical deployment scenarios for personal-use BCIs: as real BCI users heavily use their devices, a good amount of recent historical data would be readily available \cite{Card2025PersonalUse}, implying that good zero-shot performance can be expected for new sessions that are close in time.

While the previous evaluation assumes access to all available calibration sentences from the target session (typically 30--60 minutes) for model finetuning, we want to characterize \fettuccine's finetuning data efficiency. Efficient finetuning is a critical factor for real-world deployment to help ensure that users spend as little time collecting calibration data as possible. Towards this end, we ask how much calibration data (in minutes) would be required to adapt the pretrained model to each held-out session (Figure \ref{figure:wer_ft_scaling}B,E). Pretrained models achieve reasonable WERs with just 15 minutes of calibration data on average ($\sim$50 sentences), or even as few as 7 minutes for some sessions. This data requirement can be reduced further by adapting previous-session decoders instead of the pretrained model (Table \ref{tab:finetuning-schemes}).

Finally, we ask how performance on held-out sessions improves as the pretraining dataset grows, with model size held fixed to ensure portability. We construct nested pretraining subsets of $N=$ 1, 5, 10, 20, and 34 sessions by taking the $N$ sessions immediately preceding the first held-out date (05.10.2024) and walking backward, so that dataset size varies but recency is held roughly fixed across conditions. Each of these pretrained models is then finetuned on each held-out session. We find that WER improves with pretraining dataset size in an approximately log-linear fashion (Figure \ref{figure:wer_ft_scaling}C,F). Crucially, these gains are free at inference time because model size remains fixed throughout this sweep. Our results suggest that further increasing the pretraining dataset size could further reduce WERs without incurring any additional cost at inference time.

\subsection{\fettuccine provides novel ways to interpret neural data} \label{results:interpretability}
Given \fettuccine's ability to achieve high-performance text decoding from neural features, we next ask if its end-to-end architecture might enable a new type of analysis that is difficult to perform with cascaded pipelines. Because gradients in the model flow from output tokens all the way back to neural features, we hypothesize that the trained variant of \fettuccine could itself serve as a data analysis tool to reveal the spatial and temporal neural features that most significantly influenced each predicted token in the decoded output. To test this hypothesis, we develop spatial and temporal saliency map analyses based on previous methods from the computer vision literature \cite{simonyan2014deepinsidecnns, GradCAM}.

\begin{figure}[htbp]
  \centering
  \includegraphics[width=\linewidth]{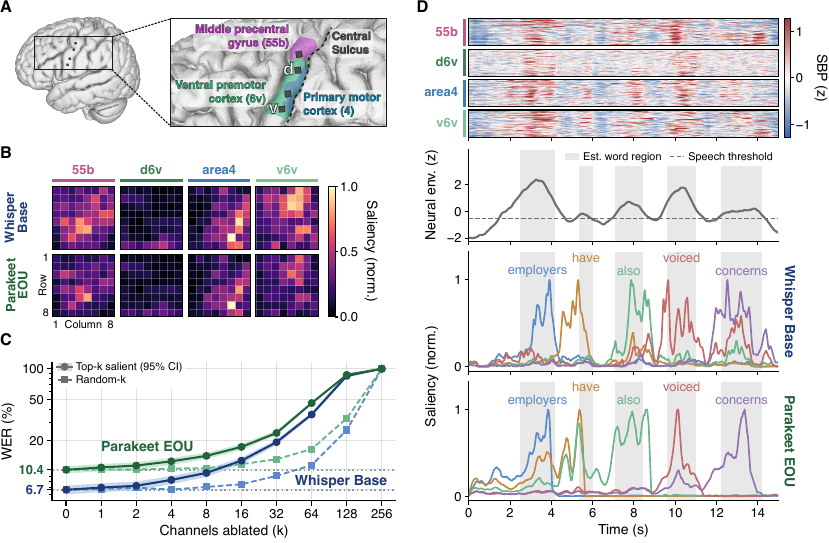}
  \caption{\textbf{Tracing neural features across space and time in end-to-end decoding.} \textbf{(A)} T15 Implant locations (adapted with permission from Card et al. \cite{Card2023SpeechNeuroprosthesisPreprint}). \textbf{(B)} Mean spatial saliency for Whisper Base (top) and Parakeet EOU (bottom) variants of \fettuccine over arrays, averaged over time and decoded tokens. Both ASR backbones independently localize importance to canonical speech motor regions and rank channels almost identically (Spearman's $\rho = 0.94$). \textbf{(C)} WER versus number of ablated channels $k$ ($\log_2$ axis). For both ASR backbones, ablating the top-$k$ most salient channels degrades WER far more steeply than a random-$k$ subset. \textbf{(D)} Per-word temporal saliency for an example sentence, aligned with that sentence's SBP, neural envelope, and estimated word boundaries. Temporal saliency maps are computed by averaging gradient saliency across channels and tokens within each word, Gaussian-smoothed, and min-max normalized independently per word.}  \label{figure:interpretability}
\end{figure}

We first test whether a spatial saliency analysis could reveal the importance of different electrodes in decoding tokens. This saliency analysis produces a spatiotemporal saliency map for each token from each validation set sentence, which we first average across tokens and then across sentences to produce spatial saliency maps (Figure \ref{figure:interpretability}B). Both Whisper Base and Parakeet EOU variants of \fettuccine independently converge onto highly similar spatial maps, with particular saliency attributed to a subset of electrodes in areas 4, v6v, and 55b. We find strong quantitative agreement in electrode importance across the spatial maps from each \fettuccine variant (Spearman's $\rho = 0.941$; $p < 0.001$), with the two models having 54 of the top 64 most salient electrodes in common. To validate the saliency maps, we ablate electrodes by saliency (Figure \ref{figure:interpretability}C). We first rank each electrode by the maximum saliency across its threshold crossing (TX) and spiking-band power (SBP) features, zero both features for the top-$k$ ranked electrodes in the raw input, and re-decode. Removing the top $k$ most salient electrodes (sweeping $k$ up to all 256 electrodes) sharply degrades WER, while removing a random set of $k$ electrodes as a control has very little effect: at $k=64$, ablating top vs. random electrodes yielded a mean per-sentence WER of 35.9\% vs. 11.6\% for Whisper Base, and 46.4\% vs. 16.6\% for Parakeet EOU, confirming the electrodes identified by the saliency maps are in fact the most important for accurately decoding the neural activity. Similarly, the saliency analyses indicate that spiking-band power features are more salient than threshold crossings across arrays, consistent with previous analyses of intracortical speech data \cite{willett2023highperformance}.

We also note that a previous coarse analysis of these data established the importance of each of the four electrode arrays by manually omitting each array from the decoding pipeline (\cite{Card2024SpeechNeuroprosthesis}, supplement), finding that the v6v, 55b, and area 4 arrays contributed the most to decoding, and d6v contributed the least. Our spatial saliency maps exhibit the same coarse pattern, but allow resolution of feature importance at a much finer scale.

We next ask if a temporal saliency analysis could reveal which time periods of neural activity are useful for decoding individual words. For a given sentence, we average the per-token spatiotemporal saliency maps across all channels, resulting in a per-token saliency timeseries. We then average the timeseries for each word's constituent tokens, Gaussian smooth, and normalize the resulting output. We find striking structure in the resulting per-word temporal saliency maps (Figure \ref{figure:interpretability}D). In particular, the ordering of the word peaks was consistent with the spoken sentence and aligned with the peaks in the neural envelope. As with the spatial maps, the temporal saliency maps from Whisper Base and Parakeet EOU variants of \fettuccine are highly similar. These temporally-localized maps confirm that \fettuccine models do not simply hallucinate output based on any implicit language modeling, but instead provide predictions grounded in the neural data. Moreover, temporal saliency maps represent a powerful tool for interrogating neural data to find the timing of different words, which may be particularly useful in helping identify word timing for paralyzed individuals that may not exhibit overt movements.

\section{Limitations}\label{limitations}
Our approach has a few notable limitations and opportunities for improvement. First, it is clear that the word error rates, while competitive amongst end-to-end models, are not the lowest observed on the brain-to-text benchmark (e.g., the top 5 submissions achieve $\leq$2\% WER). In settings where maximal accuracy and precision of language is critical (e.g., professional writing, medical discussion, legal documents), it is reasonable to opt for an expensive and bulky system if it gives better precision guarantees. Another limitation is the dependence on the availability of previously recorded sessions. When a substantial amount of data is needed (e.g., tens of sessions), this amounts to months of data collection. While not unique to \fettuccine, this is practically significant: if a model requires roughly 5--10 sessions to achieve decoding with usable word error rates as shown in Figure \ref{figure:wer_ft_scaling}, deployment benefits would be delayed accordingly. Future work could also explore the impact of more generic pretraining or transferring decoders pretrained on data from other participants. On the architecture side, all current variants of \fettuccine are not causal and can attend across the entire input window, allowing acausal token prediction. These are not deal-breakers for deployment, but they do change the way this system would be used in practice: with an acausal decoder, predictions arrive only after the entire input sequence has been processed. Therefore, while our memory gains translate directly, capitalizing on our model's latency gains will require exploring causal or small context window backbones for \fettuccine. Finally, our mobile decoding experiments explored a relatively limited set of optimizations, primarily relying on the model's direct portability after conversion to Core ML \cite{coremltools}. Future work can determine the extent to which portable brain-to-text can be achieved by further engineering and optimization of larger models. Additionally, our approach for getting models on device was fairly generic (quantization, staged load and release, etc.) and can in principle be applied to any method. We therefore encourage future work to explore the performance and tradeoffs of their methods on-device as we have here for FETTUCCINE.

\section{Discussion}
Here we present \fettuccine, a framework for adapting ASR models for brain-to-text. \fettuccine models achieve usable WERs that surpass current end-to-end methods, while being lean enough to run on mobile devices with low latency. \fettuccine also has promising generalization capabilities under finetuning, and even enables novel analyses of neural data. Altogether, \fettuccine opens the door for a new class of high performing, portable brain-to-text systems that are simpler to train and easier to extend. 

We find it surprising that ASR models across multiple model families and variants could be adapted for brain-to-text and achieve high performance with a lean intervention, considering that the base models were trained on an entirely different data domain. The broad success of this strategy suggests that as new ASR models are released, \fettuccine is likely to be able to leverage their benefits directly. Moreover, given that brain-to-text decoding has largely been dominated by cascaded architectures to date, the high performance of \fettuccine across models, while providing clear practical benefits, may suggest that end-to-end and ASR models are an underexplored space.

Several future directions follow from this work. Deploying FETTUCCINE in closed-loop clinical research sessions is a critical test to determine whether latency and reliability gains translate to live use. Cross-participant pretraining and transfer is also a promising direction for scaling FETTUCCINE beyond the single-participant models, as shown for T12 and T15 independently here. More broadly, accessibility could be expanded by leveraging the multilingual capacity of several ASR models, including variants of Whisper, making brain-to-text decoding in languages other than English potentially much easier than before.

\section*{Generative AI Usage Disclosure}
We used Claude to assist in extending our \fettuccine implementations, developing the Swift backend and mobile device decoding app, and editing manuscript prose. All AI-assisted code and text edits have been human-verified and tested to ensure integrity.

\begin{ack}
\paragraph{Funding.}
P.I.D. and C.A.M. were supported by the Georgia Tech/Emory NIH/NIBIB Training Program in Computational Neural-engineering (T32EB025816). N.H.K. acknowledges financial support from the Fonds de recherche du Québec (Secteur NT, grant 2009130) and computational resources provided by Mila and the Digital Research Alliance of Canada. C.P. was supported by the NIH Director’s New Innovator Award (DP2NS127291).
\paragraph{Competing interests.}
M.A. is the Chief Technology Officer at Constellation. S.D.S. and N.S.C. have patent applications related to speech BCI assigned to the Regents of the University of California, including intellectual property which has been licensed to a neurotechnology startup. S.D.S. is an inventor on intellectual property assigned to Stanford University related to brain-computer interfaces. He is an advisor to Subsense, Inductive Technologies, and Voyage Interfaces and was a consultant to Neuralink. C.P. has served as a consultant and research scientist for Meta (Reality Labs) and is a senior advisor to Inductive Technologies.
\end{ack}

\printbibliography

\appendix

\newpage

\section*{Appendix}
\section{Model Size vs. Performance Comparison with BIT}
\label{appendix: model_size_performance_BIT}
 
BIT \cite{BIT} is a state-of-the-art method for end-to-end brain-to-text decoding. We compare our Whisper variants with the best BIT single model E2E configuration (Aero1-Audio 1.5B decoder). We exclude the BIT ensembles (10 or 29 models merged by a fine-tuned GPT-3.5 or GPT-4). We find that \fettuccine variants outperform BIT, even at significantly smaller model sizes. 

\begin{table}[h]
  \centering
  \caption{Single model end-to-end comparison of \fettuccine and BIT \cite{BIT} on brain-to-text 2024 and 2025. BIT reports Qwen3 0.6B results on validation data.}
  \label{tab:bit_comparison}
  \small
  \setlength{\tabcolsep}{4pt}
  \begin{tabular}{lccccc}
    \toprule
          &        & \multicolumn{2}{c}{T12 WER \% $\downarrow$} & \multicolumn{2}{c}{T15 WER \% $\downarrow$} \\
    \cmidrule(lr){3-4} \cmidrule(lr){5-6}
    Model & Params (B) $\downarrow$ & val & test & val & test (pub) \\
    \midrule
    BIT E2E (Qwen3 0.6B)              & $>$0.60 & 20.61 & --    & 13.29 & --    \\
    BIT E2E (Aero1-Audio 1.5B)        & $>$1.50 & 16.07 & 15.67 & 9.46  & 11.06 \\
    \midrule
    FETTUCCINE Whisper Tiny           & 0.04 & 16.76 & 19.88 & 7.49 & 7.70 \\
    FETTUCCINE Whisper Base           & 0.08 & \underline{14.76} & \underline{15.56} & 6.72 & 6.61 \\
    FETTUCCINE Whisper Small          & 0.25 & 15.00 & 15.98 & \underline{5.29} & \textbf{4.39} \\
    FETTUCCINE Whisper Large v3 Turbo & 0.82 & \textbf{12.07} & \textbf{12.98} & \textbf{4.74} & \underline{4.64} \\
    \bottomrule
  \end{tabular}
\end{table}
 
\section{Per-Participant Compute Efficiency}
\label{appendix:compute_T12_T15}
 
Table~\ref{tab:compute_T12_T15} gives the compute efficiency of each FETTUCCINE variant on the test set of each participant. Table~\ref{table:SOTA_E2E_WER} reports the worse of the two values for each variant. RTFx (trial duration divided by inference time, averaged over sentences) is comparable only within a participant: T15 speaks more slowly than T12 (about 32 vs.\ 62 words per minute \cite{Card2024SpeechNeuroprosthesis, willett2023highperformance}), so its trials are about three times longer, while full-utterance inference latency stays roughly constant. 
 
\begin{table}[h]
  \centering
  \caption{Compute efficiency of FETTUCCINE variants on the test set of each participant (NVIDIA A40, batch size 1, greedy decoding). Table~\ref{table:SOTA_E2E_WER} reports the worse of the two values for each variant.}
  \label{tab:compute_T12_T15}
  \small
  \setlength{\tabcolsep}{4pt}
  \begin{tabular}{lccccc}
    \toprule
    Model & Params & Latency & RTFx & VRAM & RAM \\
          & B $\downarrow$ & ms $\downarrow$ & $\uparrow$ & MB $\downarrow$ & MB $\downarrow$ \\
    \midrule
    \multicolumn{6}{l}{\textit{T12 (brain-to-text 2024, $n = 1200$)}} \\
    \quad Whisper Tiny           & 0.04 & 36.3 $\pm$ 16.4  & 176  & 856  & 1999 \\
    \quad Whisper Base           & 0.08 & 50.2 $\pm$ 16.2  & 124  & 700  & 2029 \\
    \quad Whisper Small          & 0.25 & 98.1 $\pm$ 10.3  & 61   & 1594 & 2172 \\
    \quad Whisper Large v3 Turbo & 0.81 & 79.9 $\pm$ 24.7  & 77   & 2580 & 4090 \\
    \quad Parakeet TDT           & 0.62 & 48.9 $\pm$ 5.7   & 122  & 1876 & 2454 \\
    \quad Parakeet EOU           & 0.12 & 43.2 $\pm$ 6.7   & 139  & 828  & 2411 \\
    \quad Parakeet TDT-TRT       & 0.63 & 10.7 $\pm$ 7.1   & 577  & 1594 & 4465 \\
    \quad Parakeet EOU-TRT       & 0.13 & 13.4 $\pm$ 4.7   & 447  & 624  & 2449 \\
    \midrule
    \multicolumn{6}{l}{\textit{T15 (brain-to-text 2025, $n = 1450$)}} \\
    \quad Whisper Tiny           & 0.04 & 37.9 $\pm$ 5.6   & 483  & 594  & 2006 \\
    \quad Whisper Base           & 0.08 & 47.9 $\pm$ 4.1   & 377  & 694  & 2040 \\
    \quad Whisper Small          & 0.25 & 98.2 $\pm$ 7.8   & 185  & 1588 & 2196 \\
    \quad Whisper Large v3 Turbo & 0.82 & 72.4 $\pm$ 4.9   & 250  & 2578 & 4914 \\
    \quad Parakeet TDT           & 0.63 & 45.7 $\pm$ 2.3   & 398  & 1920 & 2017 \\
    \quad Parakeet EOU           & 0.12 & 42.2 $\pm$ 2.9   & 427  & 866  & 2065 \\
    \quad Parakeet TDT-TRT       & 0.63 & 12.9 $\pm$ 1.9   & 1379 & 1610 & 4660 \\
    \quad Parakeet EOU-TRT       & 0.12 & 19.6 $\pm$ 4.8   & 916  & 644  & 2412 \\
    \bottomrule
  \end{tabular}
\end{table}

\section{Ablations}
\label{appendix:ablations}

After ablating each part of the FETTUCCINE architecture and recipe, we see that each addition (U-Net, Encoder-only LoRA, word-level augmentations) earns its place in the pipeline by contributing to improvements in WER.

\begin{table}[h]
  \centering
  \caption{Ablations of architecture and augmentations on brain-to-text 2025.}
  \label{tab:ablations}
  \begin{tabular}{lccc}
    \toprule
    Run & WER val \% ($\downarrow$) & WER pub \% ($\downarrow$) & WER priv \% ($\downarrow$) \\
    \midrule
    FETTUCCINE (Whisper Tiny) & 7.49  & 7.70  & 7.61  \\
    \midrule
    1D convolutional adapter & 9.55  & 8.82  & 8.87  \\
    LoRA on encoder and decoder & 8.04  & 8.30  & 8.55  \\
    LoRA on decoder only  & 8.72  & 8.36  & 7.94  \\
    No LoRA & 12.90 & 13.15 & 13.25 \\
    No word warp & 7.78  & 7.67  & 7.33  \\
    No word crop & 7.84  & 8.00  & 7.60  \\
    No word gain & 8.24  & 8.94  & 8.29  \\
    No word augmentations & 9.56  & 10.27 & 10.04 \\
    \bottomrule
  \end{tabular}
\end{table}

\section{Choice of Neural Adapter}
\label{appendix:front_end_deets}

The Neural Adapter replaces the front-end convolution stack of each ASR backbone, mapping our $C=512$ neural channels (256 electrodes $\times$ 2 features) at 50~Hz into the $d_\text{model}$-dimensional embedding space expected by the encoder. We found that adapter design choices interact strongly with the downstream ASR family, and therefore use different architectures for the Whisper and Parakeet variants of \fettuccine.

\paragraph{Whisper: Convolutional U-Net.} For all Whisper variants we use a 1D temporal U-Net~\cite{UNet}, with the only per-variant change being the output channel count $n_\text{state}$ to match the backbone's $d_\text{model}$. The U-Net has four encoder levels with hidden widths $[512, 256, 128, 64]$, each consisting of two \texttt{Conv1d-GroupNorm-GELU-Dropout} blocks followed by stride-2 downsampling. A bottleneck block doubles the channel count ($64 \rightarrow 128$) before a symmetric decoder path with transposed convolutions and skip connections from each encoder level. A final convolution projects to $n_\text{state}$. Dropout is set to $0.1$ in the encoder and decoder paths and $0.25$ at the bottleneck. We also explored simpler alternatives, including stacks of plain convolutional blocks and U-Nets without skip connections; none matched the WER of the full U-Net for Whisper backbones.

\paragraph{Parakeet: Convolutional Subsampling Block.} For Parakeet variants we use a six-layer strided convolutional subsampling block that mirrors the front-end of the original Parakeet/FastConformer architecture~\cite{rekesh23_fastconformer, Parakeet}. The block achieves $8\times$ temporal subsampling via three stride-2 convolutions, interleaved with two stride-1 refinement layers and a final linear projection to $d_\text{model}$. We use a hidden width of $H = 512$ with GroupNorm, GELU, and dropout $0.3$ between layers. Unlike the Whisper U-Net, this design has no skip connections and produces a single $T/8$ embedding sequence, but yields feature statistics that match what the pretrained Parakeet encoder expects.

\section{Results with Beam-search Decoding}\label{appendix:beam}
Table~\ref{table:beam5_results} reports the decoding performance of \fettuccine variants on the brain-to-text 2025 dataset using beam-search decoding with a beam size of 5.
\begin{table}[h]
  \caption{Beam-search decoding ($k=5$) results for FETTUCCINE on the brain-to-text 2025 benchmark. Beam search consistently improves WER over greedy decoding (Table \ref{table:SOTA_E2E_WER}) at approximately 2$\times$ latency cost. All numbers measured on $n = 1450$ test sentences on NVIDIA A40 GPUs.}
  \label{table:beam5_results}
  \centering
  \small
  \setlength{\tabcolsep}{4pt}
  \begin{tabular}{lcccccc}
    \toprule
    \fettuccine variant & WER pub & WER priv & Latency & RTFx & VRAM & RAM \\
          & \% $\downarrow$ & \% $\downarrow$ & ms $\downarrow$ & $\uparrow$ & MB $\downarrow$ & MB $\downarrow$ \\
    \midrule
    Whisper Tiny       & 6.51          & 6.58          & 91.1 $\pm$ 15.6          & 200           & \textbf{614}  & \textbf{2103} \\
    Whisper Base       & 6.21          & 5.62          & 120.1 $\pm$ 19.2         & 151           & \underline{760}           & \underline{2114} \\
    Whisper Small      & \textbf{4.39} & \textbf{4.46} & 233.9 $\pm$ 34.3         & 77            & 1660          & 2241 \\
    Whisper Large      & \underline{4.48}          & \underline{4.74}          & 130.6 $\pm$ 16.9         & 138           & 2706          & 4917 \\
    Parakeet TDT       & 6.30          & 6.23          & \textbf{55.3} $\pm$ 3.9  & \textbf{325}  & 1878          & 2132 \\
    Parakeet EOU       & 9.21          & 8.69          & \underline{73.0} $\pm$ 11.2          & \underline{244}           & 804           & 2148 \\
    \bottomrule
  \end{tabular}
\end{table}

\section{Hyperparameter Settings: \fettuccine (Whisper variants) \label{appendix: HPs_fettuccine_whisper}}
All Whisper variants of \fettuccine share an identical training recipe and Neural Adapter architecture (Section~\ref{appendix:front_end_deets}); the only per-variant differences are the choice of pretrained Whisper backbone, which determines the encoder embedding dimension $d_\text{model}$ and (transitively) the U-Net's $n_\text{state}$ output channel count. Per-variant model statistics are summarized in Table~\ref{table:whisper_per_variant_hps}, and the shared training, optimization, LoRA, and augmentation hyperparameters are listed in Tables~\ref{table:whisper_shared_train_hps} and~\ref{table:shared_aug_hps}.

\begin{table}[h]
  \caption{Per-variant model statistics for \fettuccine Whisper models. Backbone, adapter, and LoRA columns report parameter counts within the corresponding sub-module of the trained \fettuccine model; total is their sum. The Neural Adapter (U-Net) keeps a constant capacity ($\sim$6M params) across variants, with only its final $1{\times}1$ projection scaling with $d_\text{model}$.}
  \label{table:whisper_per_variant_hps}
  \centering
  \small
  \begin{tabular}{lccccc}
    \toprule
    HuggingFace ID & $d_\text{model}$ & Backbone & Adapter & LoRA & Total \\
    \midrule
    \texttt{openai/whisper-tiny.en}        & 384  & 37.52M  & 6.12M & 98.3K  & 43.74M  \\
    \texttt{openai/whisper-base.en}        & 512  & 72.21M  & 6.19M & 196.6K & 78.59M  \\
    \texttt{openai/whisper-small.en}       & 768  & 240.96M & 6.32M & 589.8K & 247.87M \\
    \texttt{openai/whisper-large-v3-turbo} & 1280 & 806.75M & 6.58M & 1.84M  & 815.17M \\
    \bottomrule
  \end{tabular}
\end{table}

\begin{table}[h]
  \caption{Shared training, optimization, and LoRA hyperparameters for all Whisper variants of \fettuccine.
  \textsuperscript{a}Whisper Large excluded final 10 layers}
  \label{table:whisper_shared_train_hps}
  \centering
  \small
  \begin{tabular}{lcc}
    \toprule
    Hyperparameter & T15 & T12 \\
    \midrule
    \multicolumn{3}{l}{\textit{Training}} \\
    Max epochs                & 1000 & 1000 \\
    LoRA start epoch          & 35 & 35 \\
    Precision                 & bfloat16-mixed & bfloat16-mixed \\
    Gradient clip value       & 5.0 & 5.0 \\
    \midrule
    \multicolumn{3}{l}{\textit{LoRA configuration}} \\
    Rank $r$                  & 8 & 8 \\
    $\alpha$                  & 8 & 8 \\
    Dropout                   & 0.25 & 0.15 \\
    Target side               & encoder only & same \\
    Target layers             & all encoder blocks\textsuperscript{a} & same \\
    Target projections        & q/k/v/out\_proj (self-attn.) & same \\
    Bias                      & none & none \\
    \midrule
    \multicolumn{3}{l}{\textit{Optimizer (AdamWTwoStage)}} \\
    Warmup epochs             & 0 & 0 \\
    Base LR (Neural Adapter)  & $1 \times 10^{-3}$ & $1 \times 10^{-3}$ \\
    LoRA LR                   & $8 \times 10^{-4}$ & $8 \times 10^{-4}$\\
    Weight decay              & 0.1 & 0.05 \\
    Schedule                  & cosine, 0.5 cycles & cosine, 0.5 cycles\\
    Min LR (Neural Adapter)   & $8 \times 10^{-4}$ & $8 \times 10^{-4}$\\
    Min LR (LoRA)             & $1 \times 10^{-5}$ & $1 \times 10^{-7}$\\
    \bottomrule
  \end{tabular}
\end{table}

\section{Hyperparameter Settings: \fettuccine (Parakeet variants) \label{appendix: HPs_fettuccine_parakeet}}
The two Parakeet variants of \fettuccine share an identical training recipe and Neural Adapter architecture (Section~\ref{appendix:front_end_deets}); the only per-variant differences are the choice of pretrained Parakeet backbone, which determines the encoder embedding dimension $d_\text{model}$ and Conformer depth. Per-variant model statistics are summarized in Table~\ref{table:parakeet_per_variant_hps}, with shared training, optimization, LoRA, and augmentation settings in Tables~\ref{table:parakeet_shared_train_hps} and~\ref{table:shared_aug_hps}.

\begin{table}[h]
  \caption{Per-variant model statistics for \fettuccine Parakeet models. Backbone, adapter, and LoRA columns report parameter counts within the corresponding sub-module of the trained \fettuccine model; total is their sum. The Neural Adapter (subsampling block) keeps a constant capacity ($\sim$5--6M params) across variants, with only its final projection scaling with $d_\text{model}$.}
  \label{table:parakeet_per_variant_hps}
  \centering
  \small
  \begin{tabular}{lccccc}
    \toprule
    HuggingFace ID & $d_\text{model}$ & Backbone & Adapter & LoRA & Total \\
    \midrule
    \texttt{nvidia/parakeet-tdt-0.6b-v2}             & 1024 & 613.57M & 5.51M & 6.52M & 625.61M \\
    \texttt{nvidia/parakeet\_realtime\_eou\_120m-v1} & 512  & 112.52M & 5.25M & 2.33M & 120.11M \\
    \bottomrule
  \end{tabular}
\end{table}

\begin{table}[h]
  \caption{Shared training, optimization, and LoRA hyperparameters for all Parakeet variants of \fettuccine.}
  \label{table:parakeet_shared_train_hps}
  \centering
  \small
  \begin{tabular}{lcc}
    \toprule
    Hyperparameter & T15 & T12 \\
    \midrule
    \multicolumn{3}{l}{\textit{Training}} \\
    Max epochs                    & 500 & 500 \\
    LoRA start epoch              & 10 & 10 \\
    Augmentation start epoch      & 50 & 50 \\
    Precision                     & bfloat16-mixed & bfloat16-mixed \\
    Gradient clip value           & 1.0 & 1.0 \\
    Stochastic depth probability  & 0.3 & 0.1 \\
    \midrule
    \multicolumn{3}{l}{\textit{LoRA configuration}} \\
    Rank $r$                      & 8 & 8 \\
    $\alpha$                      & 8 & 8 \\
    Dropout                       & 0.3 & 0.1 \\
    Bias                          & none & none \\
    Encoder targets (per Conformer block) & self-attention (linear\_q/k/v/out), & same \\
                                  & feed-forward 1 \& 2 (linear1/2), & \\
                                  & convolution (pointwise\_conv1/2) & \\
    Joint network targets         & joint.enc, joint.pred, joint.joint\_net.2 & same \\
    \midrule
    \multicolumn{3}{l}{\textit{Neural Adapter (subsampling block)}} \\
    Hidden width $H$              & 512 & 512 \\
    Subsampling factor            & 8 & 8 \\
    Norm                          & GroupNorm (8 groups) & same \\
    Dropout                       & 0.3 & 0.2 \\
    \midrule
    \multicolumn{3}{l}{\textit{Optimizer (AdamWLoraSplit)}} \\
    Base LR (Neural Adapter)      & $5 \times 10^{-4}$ & $5 \times 10^{-4}$ \\
    LoRA LR                       & $2 \times 10^{-4}$ & $2 \times 10^{-4}$ \\
    Weight decay                  & 0.15 & 0.15 \\
    Betas                         & (0.9, 0.98) & (0.9, 0.98) \\
    Warmup epochs                 & 10 & 10 \\
    Schedule                      & cosine, 0.5 cycles & cosine, 0.5 cycles \\
    Min LR                        & $1 \times 10^{-5}$ & $1 \times 10^{-5}$ \\
    \bottomrule
  \end{tabular}
\end{table}

\section{Optimizer Details \label{appendix:optimizers}}
Both \fettuccine recipes use AdamW under the hood but differ in how learning-rate scheduling interacts with the LoRA staging.

\textbf{AdamWTwoStage} (Whisper variants) uses a single AdamW optimizer over all model parameters with one shared learning rate. The schedule consists of two distinct temporal stages, with LoRA parameters frozen externally via \texttt{requires\_grad} until they activate:
\begin{itemize}
\item \textbf{Stage 1} (epoch 0 to LoRA start epoch): the learning rate decays cosine-style from \texttt{base\_lr} to \texttt{min\_lr}. Only the Neural Adapter receives updates during this stage; LoRA parameters are present in the optimizer but frozen.
\item \textbf{Stage 2} (LoRA start epoch to end): LoRA parameters unfreeze, and the shared LR transitions to a fresh cosine schedule from \texttt{lora\_lr} down to \texttt{min\_lr2}. Both the adapter and the newly-active LoRA parameters now train at this common, smaller LR.
\end{itemize}
The effect is that LoRA's activation slows the entire model down to a more conservative regime, which we found stabilized joint training when the adapter had already partially converged.

\textbf{AdamWLoraSplit} (Parakeet variants) uses two parameter groups in a single AdamW optimizer: the Neural Adapter at \texttt{base\_lr} and LoRA parameters at \texttt{lora\_lr}. Both groups share a single cosine schedule with linear warmup that multiplicatively scales both LRs together, decaying to a common \texttt{min\_lr}. LoRA parameters are kept in the optimizer from the start (so optimizer state is preserved) but are frozen via \texttt{requires\_grad} until the LoRA start epoch, after which they begin updating on the same shared schedule as the adapter.

\section{Word-level Augmentation Details \label{appendix:augmentations}}

We introduce three word-level augmentations -- \textit{word warping}, \textit{word gain}, and \textit{word cropping} -- which operate over putative per-word neural intervals estimated from the data (Section~\ref{appendix:word_boundary_estimation}). All three are applied stochastically per trial, with hyperparameters listed in Table~\ref{table:shared_aug_hps}. Other augmentations in our pipeline (per-channel gain/offset, SpecAugment-style masking~\cite{park2019specaugment}, temporal jitter, white noise, and Gaussian smoothing) follow standard formulations from prior work~\cite{Card2024SpeechNeuroprosthesis}; their hyperparameters are also reported in Table~\ref{table:shared_aug_hps}.

\textbf{Word warping} stretches or compresses each putative word region by an independent factor sampled uniformly from $[0.5, 2.0]$ (T15 values), and each silence region by a factor from $[0.3, 3.0]$ (the wider silence range reflects the greater natural variability of inter-word pauses). The warp is applied via gather-based linear interpolation in time, with region boundaries jittered by a few samples per application to soften reliance on the (noisily estimated) word boundaries.

\textbf{Word gain} multiplies the activity within each word region by an independent scalar drawn from $\mathcal{N}(1, 0.6^2)$ (T15 values), leaving silence regions at unit gain. This is applied jointly with word warping in a single pass and is motivated by the loudness and vigor modulation observed in ventral precentral gyrus during natural speech~\cite{Srinivasan2026}.

\textbf{Word cropping} replaces the trial with a contiguous subset of its words and trims both the neural features and the transcript to match. The number of kept words is sampled bimodally: $70\%$ keep all, $10\%$ keep a single word, $20\%$ uniform over intermediate counts. Surrounding silences are retained.

\begin{table}[h]
  \caption{Augmentation hyperparameters for both ASR families. Whisper applies augmentations from epoch 0; Parakeet from epoch 50.}
  \label{table:shared_aug_hps}
  \centering
  \small
  \begin{tabular}{lcccc}
    \toprule
    & \multicolumn{2}{c}{Whisper} & \multicolumn{2}{c}{Parakeet} \\
    \cmidrule(lr){2-3} \cmidrule(lr){4-5}
    Augmentation hyperparameter         & T15        & T12          & T15        & T12 \\
    \midrule
    \multicolumn{5}{l}{\textit{Word-level (this work)}} \\
    Time warp probability               & 0.9        & 0.9          & 0.9        & 0.9 \\
    Word warp range                     & [0.5, 2.0] & [0.628, 1.2] & [0.5, 2.0] & [0.628, 1.2] \\
    Silence warp range                  & [0.3, 3.0] & [0.6, 2.556] & [0.3, 3.0] & [0.6, 2.556] \\
    Word crop probability               & 0.4        & 0.4          & 0.4        & 0.4 \\
    Word gain SD                        & 0.6        & 0.363        & 0.6        & 0.363 \\
    \midrule
    \multicolumn{5}{l}{\textit{Per-channel}} \\
    Channel gain SD                     & 0.4        & 0.501        & 0.4        & 0.501 \\
    Channel offset SD                   & 0.2        & 0.2          & 0.1        & 0.2 \\
    Mask channel probability            & 0.1        & 0.1          & 0.1        & 0.1 \\
    Max masked channels per mask        & 16         & 8            & 16         & 8 \\
    Max channel masks per trial         & 10         & 11           & 5          & 11 \\
    \midrule
    \multicolumn{5}{l}{\textit{Temporal / global}} \\
    Jitter steps                        & 30         & 16           & 20         & 16 \\
    Mask time probability               & 0.1        & 0.1          & 0.1        & 0.1 \\
    Max masked timesteps per mask       & 30         & 30           & 30         & 30 \\
    Max time masks per trial            & 10         & 11           & 5          & 11 \\
    White noise SD                      & 0.5        & 0.594        & 0.0        & 0.594 \\
    Random smoothing kernel SD range    & [0.5, 2.0] & [0.5, 2.0]   & [1.0, 2.0] & [0.5, 2.0] \\
    \midrule
    Augmentation start epoch            & 0          & 0            & 50         & 50 \\
    \bottomrule
  \end{tabular}
\end{table}
\subsection{Word boundary estimation \label{appendix:word_boundary_estimation}}
The three word-level augmentations require approximate per-word time intervals within each sentence. Because the participant attempted speech without producing audio, no ground-truth boundaries exist; we estimate them from the neural features.

For each trial, we use the z-scored features and compute two smoothed envelopes of the per-timestep summed activity: a Gaussian-smoothed signal ($\sigma = 10$ samples) and a wider uniform-smoothed signal (window $= 150$ samples). Their difference $g(t) - 0.8\, u(t)$ emphasizes transient activity over baseline; we take its cumulative sum $F(t)$, so the integrated activity over any interval $[s, e]$ is $F(e) - F(s)$. We additionally identify a sentence-level speech region $[t_\text{start}, t_\text{end}]$ by thresholding a third uniform-smoothed envelope.

Given per-word target durations $\{d_i\}$ from a TTS dictionary~\cite{pyttsx3} (rescaled to $35\%$ of the speech-region length), we place words by maximizing total integrated activity subject to monotonicity and a minimum inter-word gap $g_\text{min} = 30$ samples:
\begin{equation*}
  \max_{s_1 < s_2 < \cdots < s_W} \;\; \sum_{i=1}^{W} \big[F(s_i + d_i) - F(s_i)\big] \quad \text{s.t.} \quad s_{i+1} \geq s_i + d_i + g_\text{min}.
\end{equation*}
We solve this exactly with dynamic programming over a discretized grid of candidate start positions. The resulting boundaries are inexact but correlate well with peaks in the neural envelope (Figure~\ref{figure:interpretability}D).

\section{Mobile Decoding Details \label{appendix: mobile_decoding_details}}

Here we discuss further details relevant to the mobile decoding benchmarking. As mentioned in the main text, all experiments were run on an iPhone 15 Pro Max, which has the A17 Pro Chip. This phone had around 3.35 GB RAM available out of 7.65 GB total. For peak unified memory, we report peak unified memory as the sum of the process-wide kernel high-water mark (\texttt{phys\_footprint\_peak}), Neural Engine and GPU allocations, accessed from the iOS kernel accounting APIs.

\textbf{Whisper variants.} We decompose each Whisper model into three Core ML submodels: an encoder, a cross-attention prefill module, and a KV-cache decoder. These execute sequentially with explicit load/unload between stages so that only one submodel resides in memory at a time. The encoder processes the full neural feature sequence with all compute units available, allowing Core ML to optimally partition operations across CPU, GPU, and Neural Engine; for larger variants, it is further bisected into two chunks that are loaded and executed independently to stay within the device's memory budget. The cross-attention prefill module projects encoder hidden states into cross-attention keys and values in a single forward pass on CPU. The autoregressive KV-cache decoder then runs a greedy token-by-token loop, also on CPU, to avoid repeated transfers of KV-cache tensors between processors. 

\textbf{Parakeet variants.} Each Parakeet variant is decomposed into three Core ML submodels following the core sub-networks of the model: an encoder, an LSTM prediction network, and a joint network. The encoder runs with all compute units available, allowing Core ML to dispatch across CPU, GPU, and Neural Engine; for larger variants it is bisected into two chunks as with Whisper. The prediction and joint networks operate on single tokens per step in the greedy decoding loop and are forced to CPU to avoid repeated data transfers between processors. Both remain loaded in memory with the encoder, avoiding the staged load/unload cycle used for Whisper's decoder.

\section{Finetuning Hyperparameters, Schemes, and Additional Results}
\label{app:finetuning}

\paragraph{Hyperparameters.} For all experiments discussed in the main text, we finetuned pretrained \fettuccine models separately on each held-out session, i.e., single-session finetuning. All hyperparameters for finetuning were kept identical to the training hyperparameters for each backbone, except for the maximum number of epochs -- which we set to 100 -- and the LoRA start epoch, which was set to 0 since the pretrained model already incorporates LoRA for updating the backbone. The latter implicitly affects the learning rate schedule for the AdamWTwoStage optimizer, skipping the first stage and directly starting the second stage.

\paragraph{Other finetuning schemes.} While we focused on single-session finetuning of the pretrained model for our main experiments, we explored the impact of three other finetuning schemes on WERs:
\begin{enumerate}
    \item \textbf{Sequential single-session finetuning (S-SSFT):} Models are finetuned on calibration data from a new session, and the finetuned model is carried forth to future sessions (to be finetuned and used).
    \item \textbf{Multi-session finetuning (MSFT):} The pretrained model is finetuned on data from all held-out sessions until and including a new session.
    \item \textbf{Sequential multi-session finetuning (S-MSFT):} Models are finetuned on all calibration data until and including a new session, and this model is carried forth to future sessions (to be finetuned and used).
\end{enumerate}

\paragraph{Additional results.} Additional finetuning results on the speed of WER recovery during finetuning (vs. number of epochs), zero-shot WERs from using decoders finetuned to the immediately previous session, and WERs with other finetuning schemes are provided in Tables \ref{tab:finetuning-epochs}, \ref{tab:prev-day-zs}, and \ref{tab:finetuning-schemes} respectively.

\begin{table}[h]
\centering
\caption{Finetuning models on held-out sessions quickly rescues WERs close to usable levels.}
\label{tab:finetuning-epochs}
\begin{tabular}{lrrrr}
\toprule
& \multicolumn{4}{c}{WER (\%) $\downarrow$} \\
\cmidrule(lr){2-5}
Session & Zero-shot & Epoch 5 & Epoch 20 & Best \\
\midrule
\multicolumn{5}{l}{\textit{\fettuccine Whisper Base}} \\
05.10.2024 & 12.8 & 10.3 &  6.4 &  5.1 \\
06.14.2024 & 43.1 & 20.5 & 12.6 & 11.3 \\
07.19.2024 & 79.4 & 25.4 & 14.6 & 13.4 \\
07.21.2024 & 38.8 &  5.5 &  3.5 &  2.9 \\
07.28.2024 & 37.4 & 10.5 &  6.2 &  5.9 \\
\midrule
\multicolumn{5}{l}{\textit{\fettuccine Parakeet EOU}} \\
05.10.2024 & 21.2 &  8.3 & 10.9 &  8.3 \\
06.14.2024 & 43.0 & 26.5 & 18.5 & 15.2 \\
07.19.2024 & 77.7 & 31.7 & 21.1 & 14.9 \\
07.21.2024 & 49.6 &  9.3 &  9.9 &  7.0 \\
07.28.2024 & 48.2 & 15.4 & 11.5 &  8.2 \\
\bottomrule
\end{tabular}
\end{table}

\begin{table}[h]
\centering
\caption{Models finetuned on the previous session achieve better zero-shot performance than the pretrained model.}
\label{tab:prev-day-zs}
\begin{tabular}{lrr}
\toprule
& \multicolumn{2}{c}{Zero-shot WER (\%) $\downarrow$} \\
\cmidrule(lr){2-3}
Session & Pretrained & Prev.\ session model \\
\midrule
\multicolumn{3}{l}{\textit{\fettuccine Whisper Base}} \\
05.10.2024 & 12.8 & -- \\
06.14.2024 & 43.1 & 24.5 \\
07.19.2024 & 79.4 & 42.9 \\
07.21.2024 & 38.8 &  8.7 \\
07.28.2024 & 37.4 &  9.5 \\
\midrule
\multicolumn{3}{l}{\textit{\fettuccine Parakeet EOU}} \\
05.10.2024 & 21.2 & -- \\
06.14.2024 & 43.0 & 27.2 \\
07.19.2024 & 77.7 & 58.9 \\
07.21.2024 & 49.6 & 13.3 \\
07.28.2024 & 48.2 & 14.1 \\
\bottomrule
\end{tabular}
\end{table}

\begin{table}[h]
\centering
\caption{Evaluating the impact of different finetuning schemes on WERs.}
\label{tab:finetuning-schemes}
\begin{tabular}{lrrr}
\toprule
& \multicolumn{3}{c}{WER (\%) $\downarrow$} \\
\cmidrule(lr){2-4}
Session & S-SSFT & MSFT & S-MSFT \\
\midrule
\multicolumn{4}{l}{\textit{\fettuccine Whisper Base}} \\
05.10.2024 & 5.1 & 5.1 & 5.1 \\
06.14.2024 & 9.3 & 11.3 & 7.9 \\
07.19.2024 & 14.0 & 15.4 & 14.0 \\
07.21.2024 & 1.4 & 2.6 & 2.0 \\
07.28.2024 & 4.3 & 2.6 & 3.9 \\
\midrule
\multicolumn{4}{l}{\textit{\fettuccine Parakeet EOU}} \\
05.10.2024 & 8.3 & 8.3 & 8.3 \\
06.14.2024 & 15.2 & 12.6 & 12.6 \\
07.19.2024 & 14.9 & 14.9 & 12.9 \\
07.21.2024 & 5.5 & 4.1 & 6.4 \\
07.28.2024 & 6.9 & 5.2 & 3.9 \\
\bottomrule
\end{tabular}
\end{table}

\section{Saliency Computation and Methodology} \label{appendix: interpretability}

To interpret which features the models rely on for decoding, we used gradient-based attribution. For every non-special token $y_j$ generated during greedy decoding, we compute the absolute magnitude of the gradient of the predicted token's log-probability with respect to the raw, pre-projection neural features: $$ S = \left| \frac{\partial \log P(y_j \mid X, y_{<j})}{\partial X} \right| $$
This approach yields a saliency map of shape $512 \times T$, where the 512 channels correspond to the 256 electrodes (across 4 arrays) $\times$ 2 feature streams (threshold crossings, TX; and spiking-band power, SBP) and $T$ is the number of timesteps in the sentence (sampled at 50 Hz).

The exact steps required to compute this gradient $S$ depend on the choice of backbone. In particular, models in the Whisper family use encoder-decoder cross-attention for decoding, whereas Parakeet-based models use a transducer for decoding. To compute the gradient through Whisper models, we can directly evaluate the log-probability of token $y_j$ conditioned on the previously decoded prefix and the full encoder context. A single backwards pass computes the exact gradient from the target token logit, through the cross-attention layers and neural adapter U-Net, down to the raw input features. Conversely, Parakeet's RNN-T architecture complicates discrete token attribution because it naturally marginalizes over all possible alignments between the input sequence and the output tokens. Because a token's probability is theoretically distributed across multiple potential emission times, there is no single, deterministic gradient path to trace back to the inputs. To solve this, we use greedy-alignment pinning. During an initial forward generation pass, we trace the model's greedy decoding path to identify the exact encoder frame $t'$ at which it chooses to emit the target token. For the backwards pass, we bypass the full sequence marginalization entirely. We evaluate the joint network strictly at that single pinned intersection cell $(t', j)$, providing a clean, direct path to backpropagate the gradient through the bidirectional Conformer encoder and convolutional subsampling layers down to the raw input features.

\section{Broader Impacts}

This work aims to make speech neuroprostheses more accessible by reducing the computational requirements for brain-to-text decoding, potentially benefiting people with ALS, brainstem stroke, and other conditions that impair speech production. Mobile decoding in particular is highly beneficial, since it preserves user privacy by keeping neural data on device. However, brain-computer interfaces process highly sensitive personal data, and care must be taken to ensure that decoded outputs faithfully reflect user intent and that users retain agency over when and how their neural data are processed. Decoding errors could lead to miscommunication in high-stakes situations, and users and caregivers should be aware of current accuracy limitations.

\section{Compute Resources \label{appendix:compute_resources}}
All models were trained on an internal compute cluster consisting of NVIDIA A40s and H100s. We estimate that to reproduce our main results it would require 670 GPU hours. Reproducing mobile device inference on the Apple iPhone 15 Pro Max for most models would take less than 10 minutes each, and in the worst case, about an hour (Whisper Large).

\section{Open Source Packages Used \label{appendix:open_source}}
\begin{itemize}
    \item \texttt{torch} \cite{PyTorch} (BSD license)
    \item \texttt{pytorch-lightning} \cite{PyTorch_Lightning} (Apache 2.0 license)
    \item \texttt{ray.tune} \cite{ray_tune} (Apache 2.0 license)
    \item \texttt{scikit-learn} \cite{scikit-learn} (BSD license) 
    \item \texttt{hydra} \cite{Yadan2019Hydra} (MIT license)
    \item \texttt{transformers} \cite{wolf-etal-2020-transformers} (Apache 2.0 license)
    \item \texttt{coremltools} \cite{coremltools} (BSD-3-Clause license)
    \item \texttt{nemo} \cite{nemo} (Apache 2.0 license)
    \item \texttt{tensorrt} \cite{tensorrt} (Apache 2.0 license)
\end{itemize}

\newpage
\input{checklist.tex}

\end{document}

%% file: checklist.tex
\section*{NeurIPS Paper Checklist}

\begin{enumerate}

\item {\bf Claims}
    \item[] Question: Do the main claims made in the abstract and introduction accurately reflect the paper's contributions and scope?
    \item[] Answer: \answerYes{} %
    \item[] Justification: We propose a method for fast and efficient end-to-end brain-to-text decoding. We demonstrate our method's strong performance through validation on a publicly available brain-to-text benchmark and demonstrate its superior latency and lower memory usage through thorough benchmarking experiments. We also validate its portability and real-world applicability by deploying it on a mobile device, again with low inference latency and memory usage. We also validate the generalizability and interpretability of our framework through several experiments and analyses.

\item {\bf Limitations}
    \item[] Question: Does the paper discuss the limitations of the work performed by the authors?
    \item[] Answer: \answerYes{} %
    \item[] Justification: We have discussed the limitations of our work in detail in Section \ref{limitations}.

\item {\bf Theory assumptions and proofs}
    \item[] Question: For each theoretical result, does the paper provide the full set of assumptions and a complete (and correct) proof?
    \item[] Answer: \answerNA{} %
    \item[] Justification: This paper is application-driven and does not include theoretical results.

    \item {\bf Experimental result reproducibility}
    \item[] Question: Does the paper fully disclose all the information needed to reproduce the main experimental results of the paper to the extent that it affects the main claims and/or conclusions of the paper (regardless of whether the code and data are provided or not)?
    \item[] Answer: \answerYes{} %
    \item[] Justification: All hyperparameters and training details are provided in the Appendix.

\item {\bf Open access to data and code}
    \item[] Question: Does the paper provide open access to the data and code, with sufficient instructions to faithfully reproduce the main experimental results, as described in supplemental material?
    \item[] Answer: \answerNo{} %
    \item[] Justification: While the datasets used in this work are already publicly available, due to ongoing IP review, the full training codebase cannot be released at this time. However, we provide all details necessary for reproducing our results in the Appendix.

\item {\bf Experimental setting/details}
    \item[] Question: Does the paper specify all the training and test details (e.g., data splits, hyperparameters, how they were chosen, type of optimizer) necessary to understand the results?
    \item[] Answer: \answerYes{} %
    \item[] Justification: Data splits for training, validation, and testing were provided by the benchmark authors. Details on hyperparameter choices and optimizers are provided in the Appendix.

\item {\bf Experiment statistical significance}
    \item[] Question: Does the paper report error bars suitably and correctly defined or other appropriate information about the statistical significance of the experiments?
    \item[] Answer: \answerYes{} %
    \item[] Justification: Tables and figures reporting metrics including word error rates, latency, and memory usage include standard errors and error bars where applicable.

\item {\bf Experiments compute resources}
    \item[] Question: For each experiment, does the paper provide sufficient information on the computer resources (type of compute workers, memory, time of execution) needed to reproduce the experiments?
    \item[] Answer: \answerYes{} %
    \item[] Justification: We have provided information on the time taken to run our experiments, GPUs used, and mobile device specifications in the Appendix.
    
\item {\bf Code of ethics}
    \item[] Question: Does the research conducted in the paper conform, in every respect, with the NeurIPS Code of Ethics \url{https://neurips.cc/public/EthicsGuidelines}?
    \item[] Answer: \answerYes{} %
    \item[] Justification: The research conducted in this paper conforms with the NeurIPS Code of Ethics. Our research uses publicly available datasets with proper attribution and does not directly involve human subjects.

\item {\bf Broader impacts}
    \item[] Question: Does the paper discuss both potential positive societal impacts and negative societal impacts of the work performed?
    \item[] Answer: \answerYes{} %
    \item[] Justification: We have outlined both potential positive and negative societal impacts of this work in the ``Broader Impacts'' section in the Appendix.
    
\item {\bf Safeguards}
    \item[] Question: Does the paper describe safeguards that have been put in place for responsible release of data or models that have a high risk for misuse (e.g., pre-trained language models, image generators, or scraped datasets)?
    \item[] Answer: \answerNA{} %
    \item[] Justification: As BCI devices are not yet ubiquitous, we do not foresee any harms arising from this research in the near future.

\item {\bf Licenses for existing assets}
    \item[] Question: Are the creators or original owners of assets (e.g., code, data, models), used in the paper, properly credited and are the license and terms of use explicitly mentioned and properly respected?
    \item[] Answer: \answerYes{} %
    \item[] Justification: We have provided a list of open source packages that we used for our code development in the Appendix. The original authors of the datasets we used have been credited as well.

\item {\bf New assets}
    \item[] Question: Are new assets introduced in the paper well documented and is the documentation provided alongside the assets?
    \item[] Answer: \answerNA{} %
    \item[] Justification: We do not currently release any assets.

\item {\bf Crowdsourcing and research with human subjects}
    \item[] Question: For crowdsourcing experiments and research with human subjects, does the paper include the full text of instructions given to participants and screenshots, if applicable, as well as details about compensation (if any)? 
    \item[] Answer: \answerNA{} %
    \item[] Justification: The paper does not involve crowdsourcing and does not directly involve research with human subjects.

\item {\bf Institutional review board (IRB) approvals or equivalent for research with human subjects}
    \item[] Question: Does the paper describe potential risks incurred by study participants, whether such risks were disclosed to the subjects, and whether Institutional Review Board (IRB) approvals (or an equivalent approval/review based on the requirements of your country or institution) were obtained?
    \item[] Answer: \answerNA{} %
    \item[] Justification: The paper does not involve crowdsourcing and does not directly involve research with human subjects.

\item {\bf Declaration of LLM usage}
    \item[] Question: Does the paper describe the usage of LLMs if it is an important, original, or non-standard component of the core methods in this research? Note that if the LLM is used only for writing, editing, or formatting purposes and does \emph{not} impact the core methodology, scientific rigor, or originality of the research, declaration is not required.
    \item[] Answer: \answerNA{} %
    \item[] Justification: The core method development in this research does not involve LLMs as any component. LLMs were used to assist with editing text, aiding in data visualization, and some code development for this research. We have disclosed this in the section titled ``Generative AI Usage Disclosure''.

\end{enumerate}